\documentclass[11pt]{article}

\usepackage{graphicx}
\usepackage{epsfig}

\input babarsym
\def\BRbztodstdstNum {\ensuremath{(8.0 \pm 1.6\pm 1.2)\times 10^{-4}}}
\def\BRbztodstdst {\ensuremath{{\BR}(\Bztodstdst) = \BRbztodstdstNum}}

\def\fsideVal {1.72}
\def\fsideErr {0.10}

\def\DelELowSide {50}
\def\DelEHiSide {200\mev}
\def\mesLowSide {5.20}
\def\mesMidSide {5.26\gevcc}
\def\mesHiSide  {5.29\gevcc}

\def\Dstarp     {\ensuremath{D^{*+}}\xspace}
\def\Dstarm     {\ensuremath{D^{*-}}\xspace}
\def\Dp         {\ensuremath{D^+}\xspace}
\def\Dm         {\ensuremath{D^-}\xspace}

\def\BtoDDbar	{\ensuremath{B \to D^{(*)} \Dbar^{(*)}}\xspace}

\def\BztoDDbar	{\ensuremath{B^0 \to D^{(*)+} D^{(*)-}}\xspace}

\def\Dstptopip  {\ensuremath{\Dstarp \to \Dz\pip}\xspace}
\def\Dstptopiz  {\ensuremath{\Dstarp \to \Dp\piz}\xspace}

\def\DeltaE     {\ensuremath{\Delta E} \xspace}
\def\SsqovSpB   {\ensuremath{S^2/(S+B)}\xspace}

\newcommand{\BABARPubYear}    {01}

\newcommand{\BABARProcNumber} {74}
\newcommand{\SLACPubNumber} {9020}
\newcommand{\LANLNumber} {0000}

\def\Bztoddk   {\ensuremath{B \rightarrow D^{(*)}\overline D^{(*)}K}}    
\def\Bztodstdstzk {\ensuremath{B^0\rightarrow D^{*-}D^{(*)0}K^+}}
\def\Dpm     {\ensuremath{D^\pm}}

\newcommand{\bztoccz}{$B^0\rightarrow D^{(*)-}D^{(*)+}K^0_S$}
\newcommand{\bdstdzk}{$B^0\rightarrow D^{*-}D^{0}K^+$}
\newcommand{\bdstdstzk}{$B^0\rightarrow D^{*-}D^{*0}K^+$}

\newcommand{\btoddk}{\ensuremath{B\ra D^{(*)} D^{(*)} K}}
\newcommand{\btodsdsk}{\ensuremath{B^+ \to D^{*-} D^{*+} K^+}}
\newcommand{\de}{\ensuremath{\Delta E}}

\long\def\inst#1{\par\nobreak\kern 4pt\nobreak
    {\it #1}\par\vskip 10pt plus 3pt minus 3pt}

\begin{document}
{\pagestyle{empty}

\begin{flushright}
SLAC-PUB-\SLACPubNumber \\
\babar-PROC-\BABARPubYear/\BABARProcNumber \\
hep-ex/\LANLNumber \\
October, 2001 \\
\end{flushright}

\par\vskip 4cm

\begin{center}
\Large \bf Hadronic B decays to open charm at the \babar\ Experiment
\end{center}
\bigskip

\begin{center}
\large 
Jochen R. Schieck
\\ University of Maryland
\\ Department of Physics
\\ College Park, MD 20742
\\E-mail : {\tt schieck@slac.stanford.edu}
\\(for the \lbabar\ Collaboration)
\end{center}
\bigskip \bigskip

\begin{center}
\large \bf Abstract
\end{center}

Using about 23M $B \overline B$ events collected in 1999-2000 with the 
\babar\ detector, we report on the decays \Bztoddk\ and \Bztodstdst.
The branching fractions of the low background decay modes \Bztodstdstzk\ are determined to be 
$ {\cal B}(B^0 \rightarrow  D^{*-}D^{0}K^+) = (2.8 \pm 0.7 \pm 0.5)\times 10^{-3} $ and 
$ {\cal B}(B^0 \rightarrow D^{*-}D^{*0}K^+) = (6.8 \pm 1.7 \pm 1.7)\times 10^{-3} $,
where the first error quoted is statistical and the second systematic.  
Observation of a significant number of candidates in the color-suppressed 
decay mode $B^+\rightarrow D^{*+}D^{*-}K^+$ is reported with a branching fraction 
$ {\cal B}(B^+\rightarrow D^{*+}D^{*-}K^+)= (3.4\pm 1.6\pm 0.9)\times 10^{-3}.$
Decays of the type \BtoDDbar\ can be used to provide a measurement of the parameter \stwob 
of the Unitarity Triangle. For this decay mode we measure a branching fraction of 
\BRbztodstdst. All results presented here are preliminary.

\vfill
\begin{center}
Contributed to the Proceedings of the International Europhysics Conference on High Energy Physics,
7/12/2001--7/18/2001, Budapest, Hungary
\end{center}
\vspace{1.0cm}
\begin{center}
{\em Stanford Linear Accelerator Center, Stanford University, 
Stanford, CA 94309} \\ \vspace{0.1cm}\hrule\vspace{0.1cm}
Work supported in part by Department of Energy contract DE-AC03-76SF00515.
\end{center}

\section{Introduction}

Decays of $B$ mesons that include a charmed and an anti-charmed meson
are expected to occur through the $b$ to $c$ quark transitions 
$\overline b \to \overline c W^+$, where the $W^+$ materializes as a $c \overline s$ pair. 
These transitions are responsible for most of the $D_s$ production in $B$ 
decays. $D_s$ production has been thoroughly 
studied in experiments running at the $\Upsilon(4S)$ resonance
\cite{ref:dsdargus,ref:dsdcleo,ref:dsdcleo2}. The inclusive rate for 
 $D_s$ production in $B$ decays was recently measured by \babar, where 
a preliminary branching fraction
$ {\cal B} (B \rightarrow D_s X) = {\mathrm (10.93 \pm 
0.19 \pm 0.58 \pm 2.73)}  \times 10^{-2} $, 
the third error being due to the $D_s^+\to \Phi \pi^+$ branching fraction uncertainty,
is reported~\cite{ref:dsdbabar}.
\par Until 1994, it was believed that the $c \overline s$ quarks
would hadronize dominantly as $D_s^{+(*)}$ mesons. Therefore, 
the  branching fraction $b \to c \overline c s$ 
was computed from the inclusive $B \to D_s \, X $, 
$B \to (c \overline c)\, X $ and $B \to \Xi_c \, X $ branching fractions, leading 
 to ${\cal B}(b \rightarrow c \overline c s)={\mathrm 15.8 \pm 2.8 \%}$ 
\cite{ref:browder2}.
Theoretical calculations are unable to simultaneously describe this low 
branching fraction and the semileptonic branching fraction of the $B$ 
meson \cite{ref:bigi}. It has been conjectured 
\cite{ref:buchalla} that ${\cal B}(b \to c \overline c s)$ is in fact 
larger and that decays $B \to D \overline D K\,(X)$ 
(where D can be either a $D^0$ or a \Dpm)   
 could contribute significantly. This might also include possible decays to  
orbitally-excited $D_s$ mesons,  $B \to \overline D^{(*)} D_s^{**}$, 
followed by $D_s^{**}\to D^{(*)}\, K$. 

\par In \babar, the high statistics available allow comprehensive 
investigations to be made of the $b \to c \overline c s$ transitions. In the 
analysis described in this paper, we present evidence for the decays  
$B\to D^{(*)} \overline D^{(*)} \KS$ 
and $B\rightarrow D^{(*)} \overline D^{(*)} K$, 
using events in which both $D$'s are completely reconstructed. 
After describing the data sample and the event 
selection, we show the $D^{(*)}\overline D^{(*)}K$ signals  for 
the sum of all $B$ submodes. The branching fractions for some of the cleanest 
modes, such as $B^0 \rightarrow D^{*-} D^{(*)0} K^+$
\footnote {Charge-conjugate reactions are implied throughout this note.}, are computed.
Observation of several candidates in the 
color-suppressed decay mode  \btodsdsk\ is also reported.  
\par One of the most important goals of the \babar\ experiment is to
precisely measure the angles of the Unitarity Triangle.  While the
decay \bpsiks\ can be used to measure \stwob, the Standard
Model predicts that the time-dependent \CP violating asymmetries
in the decays \BztoDDbar can also
be used to measure the same quantity.  
An independent measurement of \stwob in
these modes would therefore provide a consistency
test of \CP-violation in the
Standard Model. The vector-vector decay \Bztodstdst\ is not, however, a pure \CP\
eigenstate.  A sizeable dilution of the measured asymmetry may be
produced by a non-negligible $P$-wave \CP-odd component.  The
dilution can, in principle, be completely removed by a time-dependent angular
analysis of the decay products~\cite{ref:dunietz}.
\par A detailed description of the \babar\ detector can be found elsewhere~\cite{ref:babar}.

\section{Analysis}
\label{sec:Analysis}

Since the $B$ mesons are produced via \epem 
$\rightarrow$ \upsbb, the energy of the $B$ in the $\Upsilon{( 4S)}$ frame 
is given by the beam energy $E_{beam}^*$, which is measured much more 
precisely than the energy of the $B$ candidate. Therefore, to isolate the 
$B$ meson signal, we use two kinematic variables: \de, the difference 
between the reconstructed energy of the $B$ candidate and the beam energy in 
the center of mass frame, and \mes, the beam energy substituted mass, 
defined as $\mes = \sqrt{E_{beam}^{*2}-p_B^{*2}}$
where $p_B^*$ is the momentum of the reconstructed $B$ in the $\Upsilon{( 4S)}$ 
frame. Signal events will have \de\ close to 0 and \mes close to the $B$ meson 
mass, 5.729\gevcc. 

\subsection{\Bztoddk\ decays}

The $B^0$ and $B^+$ mesons are reconstructed in a sample of multihadron events 
for all possible $D \overline D K$ modes, namely 
$B^0\rightarrow D^{(*)-}D^{(*)0}K^+$, 
$D^{(*)-}D^{(*)+}K^0$, $\overline D^{(*)0}D^{(*)0}K^0$ and 
$B^+\rightarrow \overline D^{(*)0}D^{(*)+}K^0$, 
$\overline D^{(*)0}D^{(*)0}K^+$, $\overline D^{(*)+}D^{(*)-}K^+$. 
The $D^0$ and $D^+$ mesons are reconstructed in the modes 
$D^0\rightarrow K^-\pi^+$, $K^-\pi^+\pi^0$, $K^-\pi^+\pi^-\pi^+$ and 
$D^+\rightarrow K^-\pi^+\pi^+$.
Charged kaon identification, 
with information from the Cherenkov angle in the DIRC and from \dedx\ 
measurements in the drift chamber and in the vertex detector, is required 
for most $D$ decay modes, as well as for the \Kpm from $B$'s.
\par $D^*$ candidates are reconstructed in the modes 
$D^{*+}\rightarrow D^0\pi^+$, $D^{*0}\rightarrow D^0\pi^0$ and    
$D^{*0}\rightarrow D^0\gamma$, by combining a $D^0$ candidate with a $\pim$, $\piz$, or photon.
Partial reconstruction of $D^{*0}$'s (no \piz or $\gamma$ 
reconstruction) is also used in the \bdstdstzk\ mode, as explained below.
$B$ candidates are reconstructed from the $D^{(*)}$, $\overline D^{(*)}$ 
and $K$ candidates. A mass constraint is applied to all the intermediate 
particles ($D^0$, $D^-$, \KS). When several candidates are selected per event in a 
specific $B$ submode ({\it e.g.} $B^+\rightarrow D^0\overline D^0 K^+$), a $\chi^2$ 
value, taking into account the difference between the measured and the PDG 
values of the $D$ masses and of the $\Delta M$ (for $D^*$'s) is constructed and 
only the candidate with the lowest $\chi^2$ value is kept for the given 
submode.  

\subsubsection{Evidence for signal in the sum of all $B$ submodes}

We present here the distributions obtained by summing all possible
\btoddk\ decay channels, for neutral and charged $B$ decays respectively.
Since multiple candidates are removed only submode by submode the same
event can appear several times in distributions obtained by summing over all modes. 
An event will appear in the peak near 0\mev\ when reconstructed
correctly, in the peak around $-160$\mev\ if it is a $D^{*}DK$
($D^*D^*K$) decay reconstructed as $DDK$ ($D^*DK$), and near the
peak around $+160$\mev\ if it is a $DDK$ ($D^*DK$) decay reconstructed
as $D^*DK$ ($D^*D^*K$).
About  120 $B^0$'s and 180 $B^\pm$ decays have been  reconstructed. 
The \mes\ distributions (Figs. 1 and 2) contain only events with $|\de|<24\mev$.
The \mes\ spectrum of $B^0$ and $B^\pm$ events can be 
fitted by the sum of a background shape and a Gaussian function used to 
extract the number of signal events. The background is empirically described 
by the ARGUS function~\cite{ref:argus}.


\begin{figure}[h]
\begin{center}
\epsfig{file=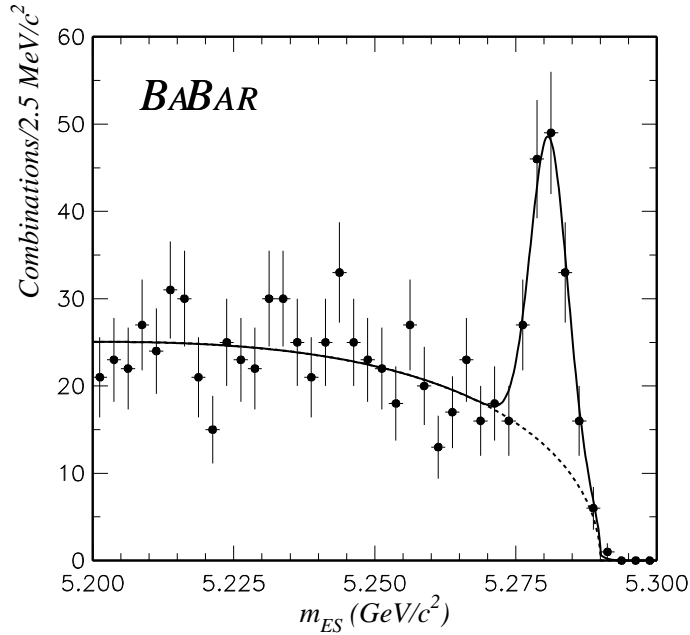,width=10cm}
\end{center}
\caption{ \mes\ distribution for the sum of all neutral modes}
\end{figure}

\begin{figure}[h]
\begin{center}
\epsfig{file=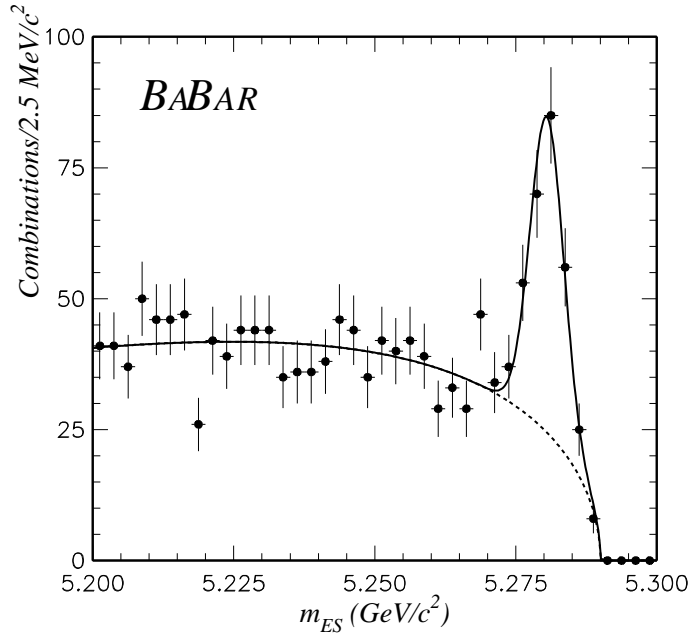,width=10cm}
\end{center}
\caption{ \mes\ distribution for the sum of charged neutral modes} 
\end{figure}

\subsubsection{Measurement of exclusive branching fractions}
In this section, we present measurements of branching fractions for the three decay
channels \bdstdzk, \bdstdstzk\ and \btodsdsk. Several candidates are also observed
in the $CP$ eigenstate \bztoccz\ but without extracting  branching fractions.
\par The selection efficiencies for each mode were obtained from detailed 
Monte Carlo simulation ~\cite{ref:geant}. 
Typical efficiencies range from 10\%, for \bdstdzk\ with both $D^0$'s 
decaying to $K\pi$,  to less than 1\%, for \btodsdsk\ with $D^0$'s 
decaying to $K\pi\pi^0$ or $K3\pi$.
Systematic errors account for the uncertainties on tracking and $\pi^0$ 
reconstruction efficiencies, $K$ identification efficiency, $D$ and $B$ 
vertexing requirements, efficiency of the requirement on $\Delta E$ used to 
define the signal box, efficiency of the $D$ mass requirement, uncertainty on 
the background shape, uncertainties on the $D$ and $D^*$ branching 
fractions, uncertainties  on the selection efficiencies arising from
Monte Carlo statistics, and uncertainty on the number of produced $B \overline B$ events
in the data sample. 
For the decay \btodsdsk a fit is performed 
with the sum of a Gaussian  function for the signal and an ARGUS function for the background. 
The number of signal events is $8.2\pm 3.5$ and 
the number of background events given by the ARGUS function is 1.7. The 
probability that the signal arises from a background fluctuation is 
$1.4\times 10^{-5}$ ($>5\sigma$). The corresponding preliminary branching fraction is 
measured to be ${\cal B}(\btodsdsk) =(3.4\pm 1.6\pm 0.9)\times 10^{-3}$
\par In the analysis of the decay $B^0\rightarrow D^{*-}D^{(*)0}K^+$  we require that 
either the  $D^0$ or the $\overline D^0$ 
decays to $K\pi$ and we do not explicitly reconstruct the $\pi^0$ or 
the photon from $D^{*0}\rightarrow D^0\pi^0$ or $D^0\gamma$.
Events containing \bdstdzk\ decays are selected by requiring $|\de|<25\mev$. 
The number of signal events is found 
to be $29.6 \pm 7.2 $. After correcting for the selection efficiencies 
and for the intermediate $D^0$ and $D^{*+}$  branching fractions 
\cite{pdg}, the preliminary branching fraction for \bdstdzk\ is found to be  
$ {\cal B}(B^0 \ra D^{*-}D^{0}K^+) = (2.8 \pm 0.7 \pm 0.5)\times 10^{-3} $.
Events containing \bdstdstzk\ decays are selected by requiring $|\de+154|<60 \mev$.
The average position and width of \de\ for \bdstdstzk\ is found to be in good 
agreement with expectations from \bdstdstzk\ signal Monte Carlo studies. 
The number of signal  events found is $80.2 \pm 15.3 $.
To extract the \bdstdstzk\ branching fraction, the contamination from 
decays $\btodsdsk$, where the $\pi^+$ from the $D^{*+}$ is not reconstructed, 
needs to be subtracted. This contribution has been estimated by performing the 
 \bdstdstzk analysis on $\btodsdsk$ signal Monte Carlo, assuming 
the $\btodsdsk$ branching fraction determined above.
The $\btodsdsk$ background contribution is  estimated to be $20.6\pm 9.7$ events.
After subtracting this contribution, the preliminary \bdstdstzk\ branching fraction is 
determined to be: 
${\cal B}(B^0 \ra D^{*-}D^{*0}K^+) = (6.8 \pm 1.7 \pm 1.7)\times 10^{-3} $.
\par As pointed out in \cite{ref:browder3}, 
the channel $B^0\rightarrow D^{*+}D^{*-}K^0_S$ is a \CP\ eigenstate that could 
be used for  $\sin 2\beta$ measurements.
For decay channels \bztoccz\ involving $D^0$'s, at 
least one decay $D^0\rightarrow K\pi$ was required. The fitted number of 
signal events is $10.1\pm 3.7$ with an estimated background of 3.4 events. 
The probability that the signal is a fluctuation
of the background is $1.4\times 10^{-5}$ ($>5\sigma$). Most of the signal is due 
to the channels $B^0\rightarrow D^{*+}D^-K^0_S$ ($4.7\pm 2.2$ events 
with a background of 1 event) and $B^0\rightarrow D^{*+}D^{*-}K^0_S$ ($4.8\pm 2.2$ events 
with a background of 0.3 event).  

\subsection{\Bztodstdst\ decay}
\Bz mesons are exclusively reconstructed by combining two charged \Dstar
candidates reconstructed in a number of \Dstar and $D$ decay modes.
The decay modes of the \Dz and \Dp used in 
this analysis were selected
by an optimization of \SsqovSpB based on Monte Carlo simulations,
where $S$ is the expected number of signal events and $B$ is the
expected number of background events.  
\Dz and \Dp candidates are reconstructed in the modes
$\Dz \to \Km \pip$, $\Dz \to \Km \pip \piz$, $\Dz \to \Km \pip \pip \pim$, $\Dz \to \KS \pip \pim$,
$\Dp \to \Km \pip \pip$, $\Dp \to \KS \pip$ and $\Dp \to \Km \Kp \pip$.   
The \Dstarp mesons are reconstructed in their decays \Dstptopip and \Dstptopiz. 
We include for this analysis the decay combinations \Dstarp\Dstarm decaying 
to (\Dz\pip,\ \Dzb\pim) or (\Dz\pip,\ \Dm\piz), but not (\Dp\piz,\
\Dm\piz) due to 
the smaller branching fraction and larger expected backgrounds.
\Dz and \Dp candidates are subjected to a mass-constraint fit and
then combined with soft pion candidates.  
Charged kaon candidates are required to be inconsistent with the pion
hypothesis, as inferred from the Cherenkov ring
measured by the DIRC and the
\dedx as measured by the SVT and DCH.  
To select \Bz candidates with well reconstructed \Dstar and $D$
mesons, we construct a $\chi^2$ that includes all PDG and measured \Dstar
and $D$ masses.
In events with more than one \Bz candidate, we chose the
candidate with the lowest value of $\chi^2$.
\par The signal region in the \DeltaE {\it vs.} \mes plane is defined to be
$|\DeltaE| < 25\mev$ and $5.273 < \mes < 5.285\gevcc$.
The width of this region corresponds to approximately $\pm 2.5\sigma$
in both \DeltaE and \mes. In addition only events with a $\chi^2 < 20 $ 
are selected.
To estimate the contribution from background in the signal region, we
define a sideband in the \DeltaE {\it vs.} \mes plane as
$ |\DeltaE| < \DelEHiSide $,
$ \mesLowSide < \mes < \mesMidSide $
and
$ \DelELowSide < |\DeltaE| < \DelEHiSide $,
$ 5.26 < \mes < \mesHiSide $. 
We parameterize the shape of the background in the \DeltaE {\it vs.}
\mes plane as the product of an ARGUS
function in \mes and a first-order polynomial in \DeltaE~.
Based on this parameterization we estimate that the ratio
of the number of background events in the signal region
to the number in the sideband region is
$(\fsideVal \pm \fsideErr)\times 10^{-2}$.
Figure 3 shows the events in the \DeltaE {\it vs.}
\mes plane after all selection criteria have been applied.  The small
box in the figure indicates the signal region defined above, and the
sideband is the entire plane excluding the region bounded by the
larger box outside the signal region.  There are a total of 38 events
located in the signal region, with 363 events in the sideband region.
The latter, together with the effective ratio of areas of the signal
region to the sideband region, implies an expected number of
background events in the signal region of $6.24 \pm 0.33 \pm 0.36$.
The quoted systematic uncertainty comes from the
background shape variation discussed previously.  Figure 4
shows a projection of the data on to the \mes axis after requiring
$|\DeltaE| < 25\mev$.
\par We use a detailed Monte Carlo simulation of the \babar\ detector to
determine the efficiency for reconstructing the signal~\cite{ref:geant}.  This,
together with the total number of \BB pairs produced during data
collection, allows us to determine a preliminary branching fraction for
\Bztodstdst to be $\BRbztodstdst$.


\begin{figure}[h]
\begin{center}
\epsfig{file=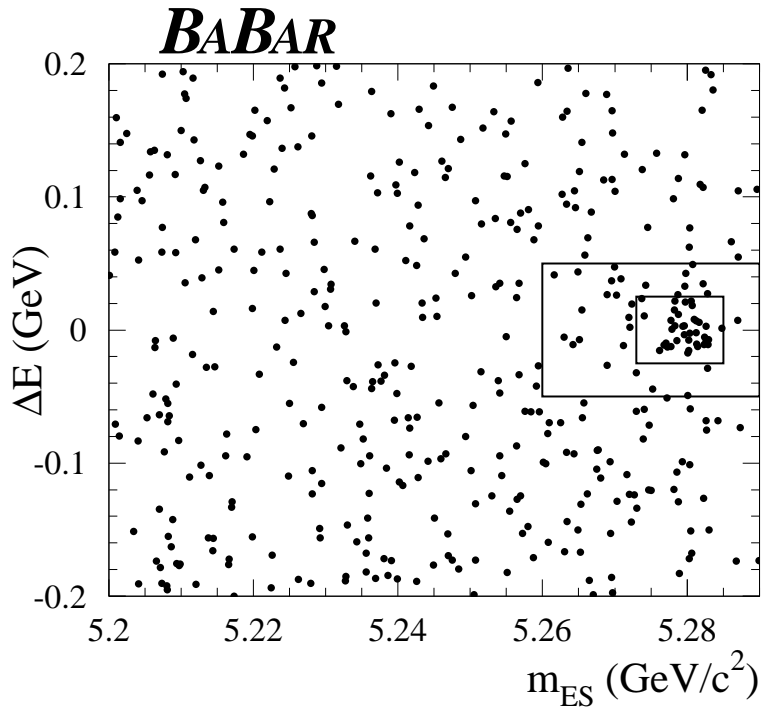,width=10cm}
\end{center}
\caption{Distribution of events in the \DeltaE {\it vs.} \mes plane.
The small box indicates the signal region, while the sideband region
is everything outside the larger box.}
\end{figure}

\begin{figure}[h]
\begin{center}
\epsfig{file=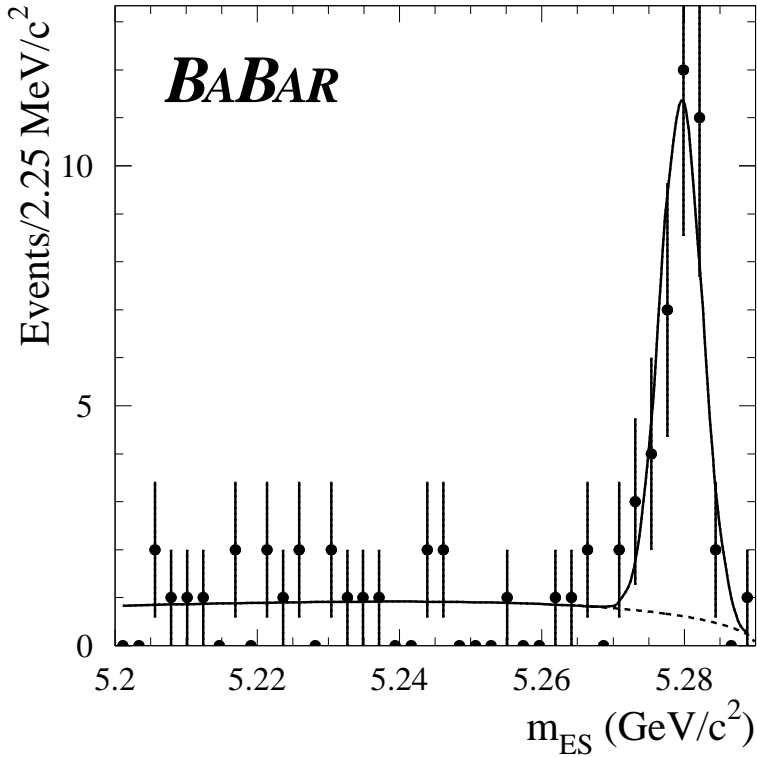,width=10cm}
\end{center}
\caption{
Distribution of events in \mes plane with a cut of $|\DeltaE|
< 25\,\mev$ applied.  The curve represents a fit to the distribution
of the sum of a Gaussian to model the signal and an ARGUS 
function to model the background shape.
}
\end{figure}

\par The dominant systematic uncertainty in this measurement comes from our
level of understanding of the charged particle tracking efficiency (9.4\%).
The high charged particle multiplicity in this decay mode makes this
measurement particularly sensitive to tracking efficiency. Uncertainties 
were assigned on a per track basis for $\pi$, $K$ and slow $\pi$,
 and were added linearly due to large correlations.  The imprecisely
known partial-wave content of the \Bztodstdst final state
is another source of systematic uncertainty (6.6\%).  This was estimated by
calculating the change in the reconstruction efficiency for different 
final angular states in Monte Carlo.
Other significant systematic uncertainties arise due to the
uncertainties on the ${\Dstar}^+$, \Dz and
$D^+$ branching fractions (5.6\%) and the differences in mass
resolutions between Monte Carlo and data (4.1\%). 
 The total systematic uncertainty from all sources is 14.5\%. 

\section{Summary}
\label{sec:Summary}
Using about 23M $B\overline B$ events, we have observed several hundred completely 
reconstructed $B\rightarrow D^{(*)}\overline D^{(*)}K$ decays. The following preliminary branching 
fractions have been measured:  
$ {\cal B}(B^0 \ra D^{*-}D^{0}K^+) = (2.8 \pm 0.7 \pm 0.5)\times 10^{-3} $ and
$ {\cal B}(B^0 \ra D^{*-}D^{*0}K^+) = (6.8 \pm 1.7 \pm 1.7)\times 10^{-3} $
in good agreement with the CLEO measurements ${\cal B}(B^0 \ra D^{*-}D^{0}K^+) 
= (4.5^{+2.5}_{-1.9} \pm 0.8)\times 10^{-3}$ and $ {\cal B}(B^0 \ra D^{*-}D^{*0}K^+) 
= (13.0^{+7.8}_{-5.8} \pm 3.6 )\times 10^{-3} $ \cite{ref:cleoddk}. 
We have observed an excess of $8.2\pm 3.5$ events over a background of 1.7 events in 
the color-suppressed decay mode $B^+\rightarrow D^{*+}D^{*-}K^+$, where
no significant number of candidates has been previously seen.
The corresponding 
preliminary branching fraction is measured to be 
$ {\cal B}(\btodsdsk)= (3.4\pm 1.6\pm 0.9) \times 10^{-3}$. 
Finally, several candidates have also been observed in the \CP\ eigenstate
$B^0\rightarrow D^{(*)+}D^{(*)-}K^0_S$.
This study confirms that the transitions  $b \rightarrow c \overline c s$ can 
proceed through the decays  $B\rightarrow D^{(*)}\overline D^{(*)}K$. 
We have observed a signal of $31.8 \pm 6.2 \pm 0.4$ events in
the decay \Bztodstdst.  We measure a preliminary branching ratio to be
$\BRbztodstdst$ in good agreement with the CLEO measurements ${\cal B}
(\Bztodstdst)=(9.9^{+4.2}_{-3.3}\pm 1.2)\times 10^{-4}  $
\cite{cleoprd62}.

\end{document}